\documentclass[aps,pra,preprint,groupedaddress]{revtex4-1}
\usepackage{color}
\usepackage{dcolumn}
\usepackage{graphicx}
\usepackage{amsmath}
\usepackage{amssymb}
\usepackage{bm}

\begin{document}


\title{Optically bistable driven-dissipative Bose-Hubbard dimer: Gutzwiller approaches and entanglement}


\author{Wim Casteels}
\affiliation{TQC, Universiteit Antwerpen, Universiteitsplein 1,
B-2610 Antwerpen, Belgium} 
\author{Michiel Wouters}
\affiliation{TQC, Universiteit Antwerpen, Universiteitsplein 1,
B-2610 Antwerpen, Belgium}

\date{\today}

\begin{abstract}
We theoretically examine the driven-dissipative Bose-Hubbard dimer in the optical bistable regime. Various approximation schemes based on a Gutzwiller mean field decoupling are applied and compared. Depending on the system parameters we show that a decoupling with respect to the real space or to the reciprocal space can be more accurate. The Gutzwiller decoupling is applied both at the level of the density matrix and for the wavefunction during a quantum trajectory simulation. The latter is shown to be a more accurate approximation. A Gaussian approximation for the non-homogeneous anti-bonding mode is also explored. We also show that entanglement in this system is witnessed by squeezing in reciprocal space.
\end{abstract}


\maketitle

\section{Introduction}
In recent years the physics of lattice structures of driven-dissipative nonlinear resonators is receiving a lot of interest (see for example Refs. \cite{ RevModPhys.85.299, ANDP:ANDP201200261, 0034-4885-79-9-096001, 2016arXiv160500383H, 2016arXiv160404433N} for recent reviews). For the description of these out-of-equilibrium systems various challenges are encountered as many of the well established theoretical and numerical tools for the description of equilibrium systems are not applicable. 
Two characteristics of these systems are that the number of particles is not conserved and the system is in a mixed state.
This led to the development of new approaches that are specifically suited for these systems (see for example Refs. \cite{PhysRevB.89.245108, PhysRevX.6.021037, 1367-2630-18-9-093007}). Advanced numerical approaches have also been recently developed that are typically based on a suitable characterization of the effective Hilbert space which can be done for example with matrix product operators \cite{PhysRevLett.98.070201, PhysRevB.78.155117, PhysRevLett.114.220601, PhysRevA.92.022116} or with the corner-space renormalization method \cite{PhysRevLett.115.080604}. 

A numerical algorithm suited for the description of such a dissipative system is the quantum trajectory simulation \cite{PhysRevLett.70.2273}, also known as wavefunction Monte Carlo \cite{PhysRevLett.68.580}. This approach describes the evolution of the system by considering an external measurement of the particles leaving the system. An initially pure state of the system then remains pure which greatly reduces the numerical complexity with respect to the evolution of the density matrix. The measurement is emulated stochastically and the density matrix can be obtained by averaging over the different realizations. These so-called {\it quantum trajectories} can give additional insight in the physics of the system since they correspond to individual experimental realizations rather than the average behavior contained in the density matrix.
An intriguing property is that the typical behavior of the quantum trajectories depends on how the photons are measured while the averaged density matrix does not \cite{Haroche:993568, 2016arXiv161201849B}. 

A widely used approximation for the description of driven-dissipative lattice structures is the Gutzwiller decoupling (see for example Refs. \cite{PhysRevLett.105.015702, PhysRevA.90.023827, PhysRevLett.110.233601, PhysRevA.94.033801, PhysRevX.6.031011, 2016arXiv161100697B}). This approach neglects all spatial correlations while the on-site correlations are fully taken into account. The problem is then reduced from a linear master equation for the density matrix with a Hilbert space dimension that is exponentially large in the system size to a coupled set of nonlinear master equations, one for each lattice site. 
Originally the Gutzwiller Ansatz was developed to approximate ground state wavefunctions \cite{PhysRevLett.10.159} and it has for example been applied for a mean-field description of the superfluid to Mott insulator transition \cite{PhysRevB.44.10328, PhysRevB.45.3137}.

An interesting perspective with these driven-dissipative photonic systems is the possibility of realizing entangled states (see for example Ref. \cite{1367-2630-15-2-025015}). This is particularly exciting since entanglement is well-known to be a key resource for new quantum technologies such as quantum computation and quantum communication \cite{Nielsen:2011:QCQ:1972505}. 

An example of such a coupled photonic structure is the driven-dissipative Bose-Hubbard dimer. This system consists of two coupled driven-dissipative nonlinear modes and has been the subject of various theoretical studies (see Refs. \cite{PhysRevB.77.125324, PhysRevLett.104.183601, PhysRevA.83.021802, 2016arXiv160702578C, PhysRevA.94.063805}). It is one of the simplest systems of which the physics is the result of an interplay between hopping, interaction, driving and dissipation. 
Besides being a convenient minimal model the driven-dissipative dimer has been experimentally realized with various photonic platforms such as semiconductor microcavities \cite{PhysRevLett.105.120403, PhysRevLett.108.126403, 2013NatPh...9..275A, 2016NatCo...711887R}, photonic-crystal lasers \cite{2015NaPho...9..311H} and superconducting circuits \cite{PhysRevX.4.031043, PhysRevLett.113.110502}.

In the first part we examine various approximation schemes that are based on the Gutzwiller mean field decoupling for the driven-dissipative Bose-Hubbard dimer. The different approximations for the density matrix are compared to the numerically determined exact solution by means of the quantum fidelity. Depending on the parameters a mean-field decoupling either in real or in reciprocal space can be more accurate. We also perform a Gutzwiller decoupling of the wavefunction in combination with a quantum trajectory simulation which is found to be a better approximation with respect to decoupling the density matrix as a direct product. A further Gaussian approximation of the non-homogeneous anti-bonding mode is examined both for the wavefunction and the reduced density matrix. In the second part we establish a relation between single mode squeezing of the collective homogeneous bonding mode and entanglement between the spatially separated resonators.

\section{The optically bistable driven-dissipative Bose-Hubbard dimer \label{Model}}
\begin{figure}[t!]
  \includegraphics[scale=0.7]{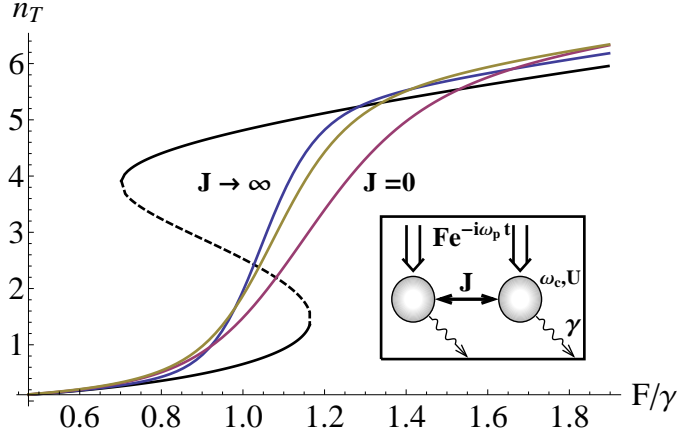}
  \caption{\label{Fig1} The total density $n_T = \langle \hat{a}_1^\dagger\hat{a}_1 \rangle + \langle \hat{a}_2^\dagger\hat{a}_2 \rangle = \langle \hat{a}_B^\dagger\hat{a}_B \rangle + \langle \hat{a}_{AB}^\dagger\hat{a}_{AB} \rangle$ of the dimer as a function of the homogeneous drive amplitude $F/\gamma$ (normalised to the loss rate $\gamma$) and for fixed $J + \Delta = 2\gamma$. The black curves are the semiclassical prediction which exhibits optical bistability with two stable branches (full lines) and one dynamically unstable branch (dashed line) (only the homogeneous solutions are presented). The results from the effective single-mode models which are exact in the limits $J \to 0$ and $J \to \infty$ are also presented. The intermediate curve is for $J = 0.25\gamma$ and is obtained from an exact diagonalisation with a truncated Hilbert space. The inset presents a schematic sketch of the dimer with the different system parameters: the hopping strength $J$, the interaction strength $U$, the cavity frequency $\omega_c$, the laser's amplitude $F$ and frequency $\omega_p$ (the laser/cavity detuning is $\Delta = \omega_p - \omega_c$) and the decay rate $\gamma$.}
\end{figure}

We discuss the dissipative Bose-Hubbard dimer with a coherent drive (see the inset of Fig. \ref{Fig1} for a sketch). We start by considering the following Hamiltonian (with $\hbar = 1$):
\begin{equation}
\hat{H} = -J\left(\hat{a}_1^\dagger\hat{a}_2+\hat{a}_2^\dagger\hat{a}_1\right) +  \sum_{j}\left(-\Delta\hat{a}_j^\dagger\hat{a}_j+\frac{U}{2}\hat{a}_j^\dagger\hat{a}_j^\dagger\hat{a}_j\hat{a}_j + F\hat{a}_j^\dagger+F^*\hat{a}_j \right),
\label{Eq: SysHam}
\end{equation} 
where  $\hat{a}^{\dagger}_j$ ($\hat{a}_j$) is the creation (destruction) operator of a boson on site $j \in\{1,2 \}$.
The first term represents the hopping between the two sites with strength $J$. The second term gives the local contributions where $\Delta = \omega_p - \omega_c$ is the laser/cavity detuning with $\omega_c$ the cavity frequency and $\omega_p$ the frequency of the coherent drive. $U$ is the interaction strength and $F$ the coherent drive amplitude which is considered to be homogeneous over the two cavities. The Hamiltonian (\ref{Eq: SysHam}) is written in the frame rotating at the drive frequency which removed the time dependence. In the quantum optical context, such a Hamiltonian can be implemented by two coupled cavity resonators with a Kerr photon-photon nonlinearity.

The Hamiltonian (\ref{Eq: SysHam}) is written in terms of the spatially separated modes, denoted as $1$ and $2$. An alternative and equivalent description can be obtained in terms of the bonding (B) and anti-bonding (AB) modes with the following annihilation operators: $\hat{a}_B = \left(\hat{a}_2 + \hat{a}_1\right)/\sqrt{2}$ and $\hat{a}_{AB} = \left(\hat{a}_2 -\hat{a}_1 \right)/\sqrt{2}$. This transforms the Hamiltonian (\ref{Eq: SysHam}) to:
\begin{eqnarray}
\hat{H} = &&\sum_{k}\left[\left(-\Delta \pm J\right)\hat{a}_k^\dagger\hat{a}_k + \frac{U}{4}\hat{a}_k^\dagger\hat{a}_k^\dagger\hat{a}_k\hat{a}_k\right] + \sqrt{2}F\left(\hat{a}_{B}^\dagger + \hat{a}_{B} \right) \nonumber\\
&&+ \frac{U}{4}\left(\hat{a}_B^\dagger\hat{a}_B^\dagger\hat{a}_{AB}\hat{a}_{AB} + \hat{a}_{AB}^\dagger\hat{a}_{AB}^\dagger\hat{a}_{B}\hat{a}_{B} + 4\hat{a}_{AB}^\dagger\hat{a}_{AB}\hat{a}_{B}^\dagger\hat{a}_{B} \right),
\label{Eq: SysHam}
\end{eqnarray} 
where the sum runs over the two reciprocal modes, i.e. $k \in \{B,AB\}$, and the linear eigenfrequency of the bonding (anti-bonding) mode is $\omega_c - J$ ($\omega_c + J$). Since we consider a homogenous drive, only the bonding mode is externally driven. The anti-bonding mode is only populated through scattering of two excitations in the bonding mode to two excitations in the anti-bonding mode. 

The losses are described within the Born-Markov approximation which results in the following Lindblad-master equation for the reduced density matrix $\hat{\rho}$ of the dimer:
\begin{equation}
i\frac{\partial\hat{\rho}}{\partial t}=\left[\hat{H},\hat{\rho}\right] + i\frac{\gamma}{2}\sum_{j}\left[2\hat{a}_j\hat{\rho}\hat{a}^\dagger_j - \hat{a}_j^\dagger\hat{a}_j\hat{\rho}-\hat{\rho}\hat{a}_j^\dagger\hat{a}_j \right],
\label{eq:Master}
\end{equation}
where $\gamma$ is the loss rate. The form of the Lindblad terms is invariant with respect to whether the spatial or reciprocal modes are considered and the sum is either over the spatial modes ($j \in \{1,2\}$) or over the bonding and the anti-bonding modes ($j \in \{B,AB\}$), depending on which description is considered.  

As a first approximation for the description of this system we consider the semiclassical or Gross-Pitaevskii approach \cite{ RevModPhys.85.299}. This corresponds to assuming the fields to be coherent, which reduces the master equation for the density matrix to two coupled differential equations for the field amplitudes $\langle \hat{a}_1\rangle$ and $\langle \hat{a}_2\rangle$. If we consider only the homogeneous solutions this leads to the following nonlinear equation for the on-site density $n = |\langle \hat{a}_1\rangle|^2 = |\langle \hat{a}_2\rangle|^2$ in the steady-state:
\begin{equation}
n = \frac{|F|^2}{\left(-\Delta -J + U n \right)^2 + \gamma^2/4},
\label{eq:SCDen}
\end{equation}
Since we only consider the homogeneous solutions, the anti-bonding mode is not occupied and all photons are in the bonding mode, i.e. $n = |\langle \hat{a}_B\rangle|^2/2$.
Eq. (\ref{eq:SCDen}) is nonlinear and admits a maximal of three solutions of which two are dynamically stable. This is known as optical bistability and can occur if $J + \Delta \geq \sqrt{3}\gamma/2$ \cite{0305-4470-13-2-034}. Eq (\ref{eq:SCDen}) depends on the detuning $\Delta$ and the hopping strength $J$ only through their sum $J + \Delta$. This means that while varying the detuning to keep the sum $J + \Delta$ constant the same result is found for any value of the hopping strength $J$. The two extreme situations are two independent cavities with $J = 0$ and an infinitely detuned anti-bonding mode with $J \to \infty$. In Fig. \ref{Fig1} the semiclassical result (\ref{eq:SCDen}) for the total density $n_T = 2n$ is presented as a function of the driving amplitude $F$ and for $J + \Delta = 2\gamma$. 
The semiclassical approach can also predict stable non-homogeneous solutions which have been observed experimentally for two spin components \cite{Paraiso:2010aa, 2016NatCo...711887R}. For sufficiently small $J$ the semiclassical approach for the driven-dissipative Bose-Hubbard dimer also predicts stable non-homogeneous solutions.

The two limits $J \to 0$ and $J \to \infty$ can be described exactly in terms of a single-mode model. The case $J = 0$ corresponds to two independent modes and for $J \to \infty$, while keeping $\Delta + J$ fixed, the anti-bonding mode is not occupied. In both cases the system is described by the driven-dissipative Kerr model for which Drummond and Walls derived an exact solution for the steady-state properties \cite{0305-4470-13-2-034} (see also appendix \ref{DW}). In Fig. \ref{Fig1} the exact results in the limits $J \to 0$ and $J \to \infty$ are also presented.
The numerically challenging regime is at intermediate values of $J$ where the single mode description breaks down. In Fig. \ref{Fig1} the result for $J = 0.25\gamma$ is also presented. This is obtained numerically by truncating the Hilbert space with a cutoff for the total number of photons and an exact diagonalisation of the master equation (\ref{eq:Master}).

\section{Gutzwiller approximation schemes}
We now apply various approximation schemes for the density matrix $\hat{\rho}$ that are based on a Gutzwiller mean field decoupling. The different approximations are compared by calculating the distance with respect to the exact density matrix $\hat{\rho}_{\text{ex}}$. This is determined through an exact diagonalisation in a truncated Hilbert space with a cutoff in the photon number. To determine the distance we consider the quantum fidelity which for density matrices $\hat{\rho}$ and $\hat{\sigma}$ is defined as: 
\begin{equation}
f(\hat{\rho},\hat{\sigma}) = \text{Tr}[\sqrt{\sqrt{\hat{\sigma}}\hat{\rho}\sqrt{\hat{\sigma}}}].
\end{equation}
The fidelity $f(\hat{\rho},\hat{\sigma})$ is symmetric: $f(\hat{\rho},\hat{\sigma}) = f(\hat{\sigma},\hat{\rho})$, it is always between $0$ and $1$ and $f(\hat{\rho},\hat{\sigma}) = 1$ if and only if $\hat{\rho} = \hat{\sigma}$ \cite{doi:10.1080/09500349414552171}. These properties motivate the use of the following distance measure between density matrices $\hat{\rho}$ and $\hat{\sigma}$:
\begin{equation}
d(\hat{\rho},\hat{\sigma}) = 1 - f(\hat{\rho},\hat{\sigma}).
\label{dist}
\end{equation}

\subsection{Real space decoupling \label{SpGW}}
The Gutzwiller decoupling is typically applied with respect to the real space degree of freedom and at the level of the density matrix \cite{PhysRevLett.105.015702, PhysRevA.90.023827, PhysRevLett.110.233601, PhysRevA.94.033801, PhysRevX.6.031011, 2016arXiv161100697B}. This approach corresponds to neglecting all spatial correlations and writing the density matrix as a direct product of single-mode density matrices $\hat{\rho}^{(i)}$ at site $i$, i.e.:
\begin{equation}
\hat{\rho} = \hat{\rho}^{(1)} \otimes \hat{\rho}^{(2)}.
\label{RDC}
\end{equation}
Since the considered set-up is homogeneous the single-mode density matrices are equal: $\hat{\rho}^{(1)} = \hat{\rho}^{(2)}$. This approach reduces the master equation (\ref{eq:Master}) to a nonlinear master equation for a single mode. Since the resulting equations are not linear they can allow multiple stable solutions (see Refs. \cite{PhysRevLett.110.233601, 2016arXiv161100697B}). 
Similar as for the semiclassical description of optical bistability this is an artifact of the approximation and the full conclusion of correlations renders these solutions metastable and leads to a unique density matrix.

\begin{figure}[t!]
  \includegraphics[scale=1]{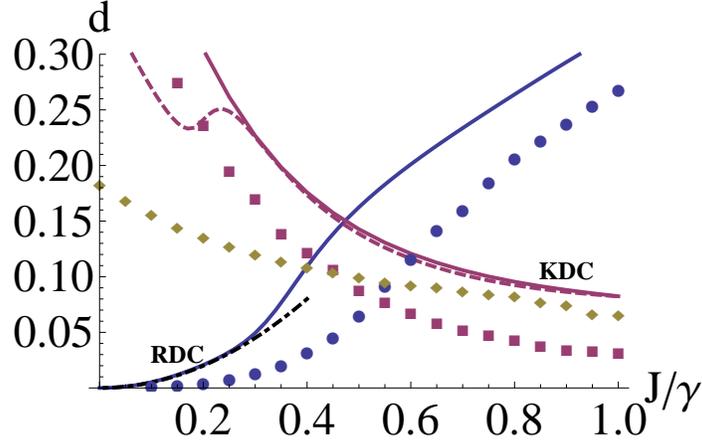}
  \caption{\label{Fig2} The distance $d$ (see Eq. (\ref{dist})) of the different approximations for the density matrix to the numerically determined exact $\hat{\rho}_{\text{ex}}$ as a function of the hopping strength $J/\gamma$. The effective detuning of the bonding mode is kept fixed at $\Delta + J = 2\gamma$ and the interaction strength is $U = \gamma$. The curves correspond to the density matrix Gutzwiller approaches and the markers are obtained from a combination of the wavefunction Gutzwiller decoupling with a quantum trajectory simulation. The results with an increasing distance as a function of $J$ correspond to a real space decoupling (RDC) while the results with a decreasing distance as a function of $J$ are the result of a reciprocal or k-space decoupling (KDC). The dashed line and the smaller squares are obtained from an additional Gaussian approximation for the anti-bonding mode. The dot-dashed line represents the power law $d = 0.5 (J/\gamma)^2$.}
\end{figure}

For the driven-dissipative Bose-Hubbard dimer this approach predicts that the density matrices $\hat{\rho}^{(i)}$ satisfy the master equation of the driven-dissipative Kerr model. This master equation is not linear since the effective drive amplitude $F^{\text{eff}}$ depends on the expectation value of the field: $F^{\text{eff}} = F - J\langle \hat{a} \rangle$. This was also found in Ref. \cite{PhysRevLett.110.233601} for the driven-dissipative Bose-Hubbard model. In Fig. \ref{Fig2} the full curve denoted as RDC (real space decoupling) denotes the distance (\ref{dist}) between the density matrix (\ref{RDC}) and the exact density matrix $\hat{\rho}_{\text{ex}}$ as a function of the hopping strength $J$. The sum $\Delta + J = 2\gamma$ is kept fixed and the interaction strength is taken $U = \gamma$. For the considered parameters the approach predicts multiple solutions for $J > 1.53$, which is not shown in Fig. \ref{Fig2}. The distance $d$ goes to zero for $J \to 0$ for which the resonators become independent. For small $J$ the distance increases according to a power law and for the considered parameters $d = 0.5 (J/\gamma)^2$ results in a good fit (see dot-dashed line in Fig. \ref{Fig2}). The distance becomes larger as the hopping strength increases since the spatial correlations become more important.

Historically the Gutzwiller Ansatz was introduced for groundstate wavefunctions \cite{PhysRevLett.10.159, PhysRevB.44.10328, PhysRevB.45.3137}. Later, this approach was extended for the description of mixed density matrices as discussed above. In both cases this is equivalent to considering an effective Hamiltonian with parameters that depend on the expectation value of system operators. The extension to a mixed density matrix might seem questionable since these expectation values then correspond to an ensemble average. After all, an individual realization does not have access to such information about the average behavior. This could become especially problematic in the optical bistable regime since the expectation values then correspond roughly to an average over two metastable states. This shows that the Gutzwiller Ansatz for a mixed density matrix is not the same approximation as a Gutzwiller decoupling of a wavefunction. 

To examine this further we have also approximated the density matrix by applying a Gutzwiller decoupling of the wavefunction in combination with a quantum trajectory simulation with a photon counting measurement (see appendix \ref{ALG}). The pure state of the system is then evolved in a truncated Hilbert space consisting of the product wavefunctions $\left| \psi_1 \right>\left| \psi_2 \right>$. The density matrix is then obtained by averaging over the density operator $\rvert \psi_1 \rangle\rvert \psi_2 \rangle \langle \psi_1\rvert \langle \psi_2\rvert$. The spatial quantum correlations are thus neglected but classical correlations are incorporated. The circles in Fig. \ref{Fig2} represent the distance of the obtained density matrix with the exact solution. This approach is again exact in the limit $J \to 0$ where there are no spatial correlations. As the hopping strength increases quantum correlations become important and the distance increases. A comparison with the density matrix Gutzwiller approach (\ref{RDC}) in Fig. \ref{Fig2} reveals that the decoupling of the wavefunction is more accurate. 

\subsection{Reciprocal space decoupling}
There is no a priori reason for applying the Gutzwiller decoupling with respect to the spatial degree of freedom. 
Moreover, since for $J \to \infty$ only the bonding mode is populated we expect a Gutzwiller decoupling with respect to the reciprocal modes to be more accurate for large hopping strength. We start by applying the Gutzwiller decoupling at the level of the density matrix $\hat{\rho}$ by writing $\hat{\rho}$ as a direct product of single mode density matrices for the bonding and the anti-bonding modes: 
\begin{equation}
\hat{\rho} = \hat{\rho}^{(B)} \otimes \hat{\rho}^{(AB)}.
\label{KDC}
\end{equation} 
The reciprocal modes are not homogeneous, i.e. $\hat{\rho}^{(B)} \neq \hat{\rho}^{(AB)}$. 
This leads to two coupled single mode master equations for $\hat{\rho}^{(B)}$ and $\hat{\rho}^{(AB)}$. The decoupling leads to effective two photon driving processes and also in this case an exact solution for the steady-state is known \cite{minganti2016exact, PhysRevA.94.033841}. In Fig. \ref{Fig2} the distance of the resulting density matrix with the exact solution $\hat{\rho}_{\text{ex}}$ is indicated by KDC (k-space decoupling). The approximation becomes better as the hopping strength is increased and becomes exact in the limit $J \to \infty$ where the anti-bonding mode is not occupied. For large $J$ we again find that the distance follows a power law and for the considered parameters $d = 0.16 (J/\gamma)^{-1.86}$ results in a good fit (not shown in the figure).

So far we have considered the full Hilbert space for the effective single mode master equations. Both for the decoupling in real space and in reciprocal space this can be done efficiently thanks to the existence of exact solutions \cite{0305-4470-13-2-034, minganti2016exact, PhysRevA.94.033841}. This is however not generally the case and a further approximation can be helpful for an efficient calculation. We do this by truncating the Hilbert space to the subspace consisting of the Gaussian or quadratic states. 
In general this approach breaks down for an optical bistable system since the bimodality is clearly not captured by a quadratic density matrix. 
This means that this approximation does not capture the switching dynamics between the metastable branches and we can not use is for the bonding mode density matrix $\hat{\rho}^{(B)}$. 
Instead we only approximate the anti-bonding mode density matrix $\hat{\rho}^{(AB)}$ by a Gaussian state which is done by applying Wick's theorem (see appendix \ref{GaussDM}). 
The dashed line in Fig. \ref{Fig2} indicates the distance with respect to the exact density matrix $\hat{\rho}_{\text{ex}}$. For large values of $J$ a good agreement is found with the result obtained by considering the full Hilbert space. For small values of $J$ where the reciprocal Gutzwiller decoupling is not so accurate the two results deviate and the distance even becomes smaller for the Gaussian approximation.

As discussed in the previous section the application of the Gutzwiller decoupling (\ref{KDC}) at the level of the density matrix can be questionable, especially in the optical bistable regime. We also combined a quantum trajectory simulation with a Gutzwiller decoupling of the wavefunction in the reciprocal space, i.e. $\left| \psi \right> = \left| \psi_B \right>\left| \psi_{AB} \right>$. The escaped photons are then counted in the reciprocal basis. The large red squares in \ref{Fig2} indicate the resulting distance with respect to the exact solution $\hat{\rho}_{\text{ex}}$. As for the real space decoupling this is more accurate with respect to neglecting all the correlations between the reciprocal modes (\ref{KDC}). 

Performing the quantum trajectory simulation for the bonding and anti-bonding modes with a product wavefunction can still be numerically expensive. A possible solution is to truncate the Hilbert space to Gaussian wavefunctions. We have done this for the wavefunction of the anti-bonding mode (see appendix \ref{GaussWF}). 
The small diamonds in Fig. \ref{Fig2} denote the distance with the exact solution $\hat{\rho}_{\text{ex}}$. If the full Hilbert space is considered a better result is found for relatively large $J$ but the Gaussian approximation is still better than the density matrix decoupling (\ref{KDC}). For small $J$, where the reciprocal decoupling is not very accurate, the distance becomes even smaller than for the result with the full single-mode Hilbert space. 
We have also tried the Gaussian ansatz for the bonding mode but this unfortunately led to qualitatively wrong predictions for the switching behavior. 


\section{single mode squeezing and two-mode entanglement}
As discussed in Section \ref{Model} only the bonding mode of the dimer is populated in the limit $J \to \infty$ while keeping fixed $\Delta + J$. Since this mode is non-local in space, an intriguing  question is how a non-classical single-mode state for the bonding mode translates to quantum correlations between the spatially separated resonators. In particular we wonder whether this can lead to entanglement corresponding to a non-separable density matrix. A separable density matrix can, by definition, be written as 
\begin{equation}
\hat{\rho} = \sum_i p_i \hat{\rho}^{(1)}_i \otimes \hat{\rho}^{(2)}_i,
\label{SEP}
\end{equation}
with $\sum_ip_i = 1$. 

To examine entanglement between two modes we consider the widely applied criterion derived by Duan and Simon \cite{PhysRevLett.84.2722, PhysRevLett.84.2726}. Two pairs of canonically conjugated operators are considered: $\{\hat{x}_i\}$ and $\{\hat{p_i}\}$, with $i \in \{1,2\}$, such that $\left [\hat{x}_i,\hat{p}_j \right] = i\delta_{i,j}$. The following two EPR-like operators can then be constructed: $\hat{u} = \hat{x}_1 + \hat{x_2}$ and $\hat{v} = \hat{p}_1 - \hat{p}_2$. If the state is separable and can be written as (\ref{SEP}) the total variance of any pair of EPR-like operators $\hat{u}$ and $\hat{v}$ satisfies the following inequality \cite{PhysRevLett.84.2722}:
\begin{equation}
(\Delta u)^2 + (\Delta v)^2 \geq 2,
\label{EntCrit}
\end{equation}
where we used the following notation for the variance of an operator $\hat{O}$: $(\Delta O)^2 = \langle \hat{O}^2 \rangle - \langle \hat{O}\rangle^2$.
A violation of the inequality (\ref{EntCrit}) is thus a sufficient criterion for entanglement.

We now apply this criterion to examine the entanglement between the two spatial modes of the driven-dissiative Bose-Hubbard dimer. We start by rewriting the operators $\hat{x}_i$ and $\hat{p_i}$ for the spatial modes in terms of the canonically conjugated operators for the bonding and the anti-bonding modes: 
\begin{eqnarray}
\hat{x}_1 &= \left(\hat{x}_B + \hat{x}_{AB}\right)/\sqrt{2}; \\
\hat{x}_2 &= \left(\hat{x}_B - \hat{x}_{AB}\right)/\sqrt{2}; \\
\hat{p_1} &= \left(\hat{p}_B + \hat{p}_{AB}\right)/\sqrt{2}; \\
\hat{p_2} &= \left(\hat{p}_B - \hat{p}_{AB}\right)/\sqrt{2}.
\end{eqnarray}
The EPR-like operators can then be written as:
\begin{eqnarray}
\hat{u} = \hat{x}_1 + \hat{x_2} = \sqrt{2}\hat{x}_B; \nonumber \\
\hat{v} = \hat{p}_1 - \hat{p}_2 =  \sqrt{2}\hat{p}_{AB}.
\label{EPR}
\end{eqnarray}
Assuming only the bonding mode to be populated and the anti-bonding mode to be in the vacuum gives:
\begin{eqnarray}
(\Delta v)^2 = 2(\Delta p_{AB})^2 = 1
\end{eqnarray}
Using the criterion for separability of Ref. \cite{PhysRevLett.84.2722} we find that the state is entangled if 
\begin{equation}
(\Delta u)^2 = 2(\Delta x_B)^2 < 1.
\end{equation}
This corresponds exactly to the single-mode squeezing condition for the bonding mode. This reveals that if the anti-bonding mode is not populated, squeezing of the bonding mode corresponds to entanglement between the two spatially separated modes.
This is particularly exciting since single mode squeezing has been experimentally realized with various experimental platforms such as optical cavities \cite{PhysRevLett.55.2409}, semiconductor microcavities \cite{boulier2014polariton, PhysRevA.69.031802} and circuit-QED \cite{PhysRevLett.60.764, castellanos2008amplification}. 

\begin{figure}[t!]
  \includegraphics[scale=0.8]{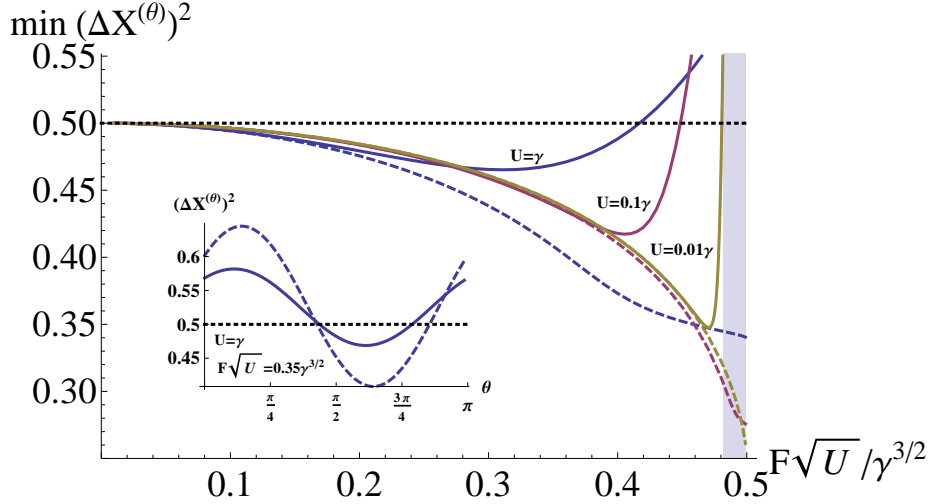}
  \caption{\label{Fig3} The minimal variance of the quadrature $\text{min}(X^{(\theta)})^2$ as a function of $F\sqrt{U}/\gamma^{3/2}$ for the driven-dissipative Kerr model with $\Delta = \gamma$. The minimization of $(X^{(\theta)})^2$ is with respect to the angle $\theta$. The full lines are the exact quantum results for the steady-state for three different values of the nonlinearity: $U/\gamma = 1$, $0.1$ and $0.01$. The dashed curves are obtained from a Gaussian approximation (results for the upper branch are not shown). The shaded area indicates the region where the semiclassical approach predicts optical bistability. The inset shows the variance as a function of $\theta$ for $U = \gamma$ and $F\sqrt{U} = 0.35\gamma^{3/2}$ together with the Gaussian approximation (dashed line).}
\end{figure}

To make this more precise we start by resuming some of the results on single mode squeezing for the driven-dissipative Kerr model. We consider the following quadrature operators for a mode with annihilation operator $\hat{a}$:
\begin{eqnarray}
\hat{X}^{(\theta)} = \frac{1}{\sqrt{2}}\left( e^{i \theta}\hat{a}^\dagger + e^{-i \theta}\hat{a}\right); \\
\hat{P}^{(\theta)} = \frac{i}{\sqrt{2}}\left( e^{i \theta}\hat{a}^\dagger - e^{-i \theta}\hat{a}\right).
\end{eqnarray}
These operators are canonically conjugate since they satisfy the commutation relation $\left [\hat{X}^{(\theta)},\hat{P}^{(\theta)} \right] = i$, for any value of the angle $\theta$.
In Fig. \ref{Fig3} the minimal variance of the quadrature $\text{min} (\Delta X^{(\theta)})^2$ is presented, where the minimization is with respect to $\theta$, with $\Delta = \gamma$ and various values for the nonlinearity $U$. If the variance of the quadrature is smaller than $1/2$ the mode is squeezed (for a coherent state the variances of all quadratures are $1/2$). The drive amplitude in Fig. \ref{Fig3} is rescaled with the square root of the nonlinearity which allows to examine the role of the nonlinearity on a single scale \cite{2016arXiv160800717C}. These results are obtained using the exact expressions for the steady-state properties of the driven-dissipative Kerr model derived in Ref. \cite{0305-4470-13-2-034} (see also appendix \ref{DW}). The shaded area in Fig. \ref{Fig3} indicates the region where the semiclassical approach predicts optical bistability. As the drive amplitude is increased the mode initially becomes increasingly squeezed. As the optical bistability region is approached the switching between the semiclassically stable branches leads to an increase of the variance and finally the variance becomes larger than $1/2$ and the mode is no longer squeezed.

The dashed lines in Fig. \ref{Fig3} are the prediction for the minimal variance $\text{min} (X^{(\theta)})^2$ obtained from a quadratic approximation of the fluctuations around the semiclassical solution (see appendix \ref{DW}). For a small driving amplitude the same qualitative behavior is found as for the exact result and for a small nonlinearity the results also agree quantitatively. As the bistability region is approached the results deviate and the quadratic prediction for the variance decreases further. This is a consequence of the fact that this approximation does not capture the switching between the metastable branches. 
From an experimental point of view a comparison of the timescales determines which prediction can be observed \cite{2016arXiv160800260R}. If the switching timescale is much longer than the experimental timescale there is no time for the system to explore the other metastable solution and the quadratic approximation is better. This is the case for a small nonlinearity $U$ and/or a large detuning $\Delta$. 

\begin{figure}[t!]
  \includegraphics[scale=0.4]{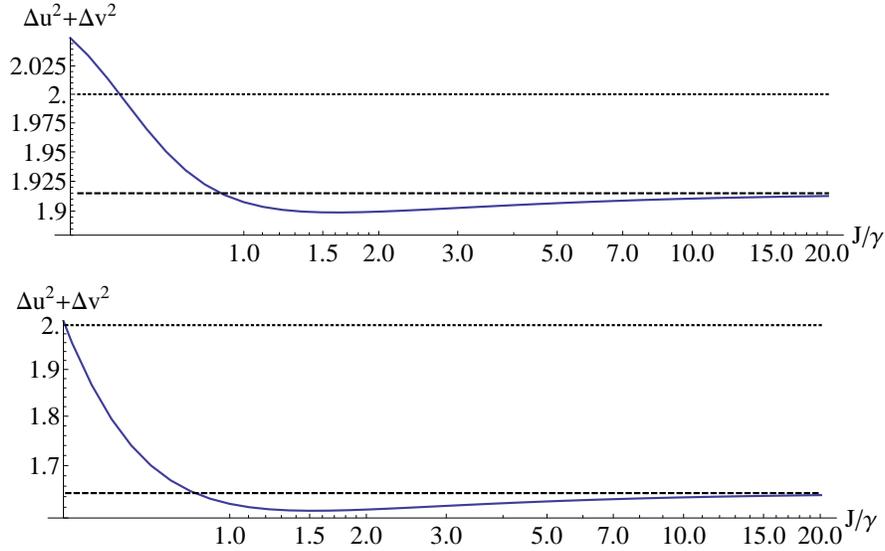}
  \caption{\label{Fig4} The sum of the variances of the EPR-like operators (\ref{EPR}) as a function of the hopping strength $J/\gamma$ (in units of $\gamma$ on a logarithmic scale) with a detuning $\Delta = \gamma - J$. The upper panel is the exact steady-state result for $U = \gamma$ and $F\sqrt{U} = 0.33\gamma^{3/2}$ and the lower panel is the Gaussian approximation for $U = 0.01\gamma$ and $F\sqrt{U} = 0.48\gamma^{3/2}$. If the sum is lower than $2$ (dotted line) the resonators are entangled. The dashed line gives the single mode approximation for an unoccupied anti-bonding mode which is valid in the limit $J \to \infty$.}
\end{figure}

We now have all the ingredients to examine the entanglement between the two spatially separated modes of the dimer. In Fig. \ref{Fig4} the sum of the variances of the EPR-like operators (\ref{EPR}) is presented as a function of the hopping strength while keeping fixed $\Delta + J = \gamma$. The angle $\theta$ for the quadratures is taken such that the sum of the variances is minimal in the limit $J \to \infty$. The upper panel in Fig. \ref{Fig4} is obtained from an exact diagonalization of the master equation (\ref{eq:Master}) for the steady state, with $U = \gamma$ and $F\sqrt{U} = 0.33\gamma^{3/2}$. The lower panel is obtained from a Gaussian approximation for the fluctuations around the semiclassical prediction with $U = 0.01\gamma$ and $F\sqrt{U} = 0.48\gamma^{3/2}$ (see appendix \ref{DDDGauss}). The dashed lines are the single mode description: $\langle(\Delta \hat{u})^2\rangle + \langle(\Delta \hat{v})^2\rangle = 2\text{min}_\theta\langle(\Delta \hat{X}^\theta_B)^2\rangle + 1$ which is valid if the anti-bonding mode is unoccupied. Indeed, the results converges to this value in the limit $J \to \infty$. According to the entanglement criterion discussed above the modes are entangled if the sum of the two variances is smaller than $2$. In Fig. \ref{Fig4} we see that only for small values of the hopping strength $J$ this criterion is not satisfied. This was expected since for $J \to 0$ the modes are not coupled and thus not entangled. From Fig. \ref{Fig4} we see that the maximal violation of the entanglement criterion is not obtained in the limit $J \to \infty$ but for a finite value of $J$. The variances of the quadratures can be experimentally measured with a standard homodyne detection scheme. 

\section{Conclusions and perspectives}
We examined various Gutzwiller mean field decoupling schemes for the description of the driven-dissipative Bose-Hubbard dimer in the optical bistable regime. This revealed that depending on the system parameters a Gutzwiller decoupling either in real or in reciprocal space can be more accurate. We explored two possibilities for the Gutzwiller decoupling: either by writing the density matrix as a direct product or by performing a quantum trajectory simulation with a product wavefunction. The latter was done with an external counting measurement of the photons and led to a better approximation for all considered parameters. We also showed that a more efficient simulation is possible if the Hilbert space of the anti-bonding mode is truncated to Gaussian states, both at the level of the density matrix and of the wavefunction. 

This leads to various exciting perspectives for the simulation of multi-mode systems. All the presented schemes can be straightforwardly extended to larger systems.  
If more modes are considered, an extension of the Gaussian truncation schemes could allow to capture some of the quantum correlations between different modes which are neglected with a Gutzwiller mean field decoupling. For example for a homogeneous driven-dissipative quantum fluid with periodic boundary conditions the modes with wavevectors $\vec{k}$ and $-\vec{k}$ are entangled \cite{PhysRevA.89.043819}. This is not captured by a Gutzwiller decoupling but can be incorporated in a Gaussian approximation by considering two-mode squeezing. 

An intriguing question concerns the way the escaping photon are measured during the quantum trajectory simulations with a product wavefunction. We only considered photon counting measurements but one could for example also consider a homodyne detection. By considering a different measurement the quantum trajectories can have less entanglement \cite{PhysRevLett.93.120408}, leading to a better result. 

We also examined the relation between squeezing of the homogeneous mode in the reciprocal space and entanglement between the spatially separated modes. Using a sufficient criterion for entanglement we found that the spatial modes are entangled if only the bonding mode is populated and if it is squeezed. This picture was confirmed with a numerical analysis. 
An interesting perspective is how this generalizes to larger systems with more spatially separated coupled modes. 

\acknowledgements{We gratefully acknowledge discussions with N. Bartolo, C. Ciuti, R. Fazio, V. Gladilin, F. Minganti, R. Rota, M. Van Regemortel and W. Verstraelen. We acknowledge support from the FWO-Odysseus program. }

\appendix 

\section{Steady-state properties of the driven-dissipative Kerr model \label{DW}}
\subsection{Exact solution}
We now summarize the exact expressions for the steady-state properties of the driven-dissipative Kerr model originally derived by Drummond and Walls in Ref. \cite{0305-4470-13-2-034}. The Hamiltonian is given by (with $\hbar = 1$):
\begin{equation}
\hat{H}_{\text{Kerr}} = -\Delta\hat{a}^\dagger\hat{a}+\frac{U}{2}\hat{a}^\dagger\hat{a}^\dagger\hat{a}\hat{a} + F\hat{a}^\dagger+F^*\hat{a},
\label{Eq: KerrHam}
\end{equation} 
with $\hat{a}^\dagger$ ($\hat{a}$) the creation (annihilation) operator for an excitation. The Hamiltonian is written in the frame rotating at the drive frequency which removed the time-dependence. The model parameters are the drive/cavity detuning $\Delta$, the interaction strength $U$ and the coherent drive amplitude $F$. The losses are described with the Born-Markov approximation which leads to the following Lindblad-master equation for the reduced density matrix $\hat{\rho}$:
\begin{equation}
i\frac{\partial\hat{\rho}}{\partial t}=\left[\hat{H}_{\text{Kerr}},\hat{\rho}\right] + i\frac{\gamma}{2}\left[2\hat{a}\hat{\rho}\hat{a}^\dagger - \hat{a}^\dagger\hat{a}\hat{\rho}-\hat{\rho}\hat{a}^\dagger\hat{a} \right],
\label{eq:KerrMaster}
\end{equation}
with $\gamma$ the loss rate. In Ref. \cite{0305-4470-13-2-034} the generalized P-representation was used to derive the following expression for the steady-state correlation functions:
\begin{equation}
\langle \hat{a}^{\dagger n}\hat{a}^m \rangle = \left(-\frac{2 F}{U} \right)^{n+m} \frac{\Gamma(c)\Gamma(c^*)}{\Gamma(c+m)\Gamma(c^*+n)}\frac{\mathcal{F}(c+m,c^*+n,8|F/U|^2)}{\mathcal{F}(c,c^*,8|F/U|^2)},
\label{eq:DWCF}
\end{equation}
with $c = 2\left( -\Delta - i\gamma/2 \right) /U$, $\Gamma(x)$ the gamma function and $\mathcal{F}(c,d,z)$ the hypergeometric function:
\begin{equation}
\mathcal{F}(c,d,z) = \sum_{n=0}^\infty \frac{\Gamma(c)\Gamma(d)}{\Gamma(c+n)\Gamma(d+n)}\frac{z^n}{n!}.
\end{equation}

The results for the density presented in Fig. \ref{Fig1} for the extreme cases $J = 0$ and $J \to \infty$ are both obtained by applying Eq. (\ref{eq:DWCF}) with $n = m = 1$. For $J = 0$ the total density is simply twice this density. For $J \to \infty$ the total density equals the density of the bonding mode which is given by Eq. (\ref{eq:DWCF}) with $n = m = 1$ with an interaction strength $U/2$ and coherent drive amplitude $\sqrt{2}F$.

The results for the variance of the quadrature $\hat{X}^{(\theta)}$ in Fig. \ref{Fig3} are also obtained from expression (\ref{eq:DWCF}). The variance can be written as:
\begin{eqnarray}
(\Delta X^{(\theta)})^2 &&= \langle \hat{X}^{(\theta)2} \rangle - \langle \hat{X}^{(\theta)}\rangle^2 \nonumber\\
 && = e^{2i \theta} \frac{\langle\hat{a}^{\dagger 2} \rangle - \langle\hat{a}^{\dagger} \rangle^2 }{2} + e^{-2i \theta}\frac{\langle \hat{a}^2 \rangle - \langle \hat{a} \rangle^2 }{2}+ \langle \hat{a}^\dagger\hat{a} \rangle - \langle \hat{a}^\dagger \rangle\langle \hat{a} \rangle +1/2.
\label{QuadVar}
\end{eqnarray}
These are all expectation values of the form (\ref{eq:DWCF}).

\subsection{Gaussian approximation}
The semiclassical steady-state prediction for the field $\alpha = \langle \hat{a} \rangle$ is determined by the following equation:
\begin{equation}
\left(-\Delta - i\frac{\gamma}{2} + U |\alpha|^2 \right)\alpha + F = 0.
\end{equation}
Assuming the fluctuations around the semiclassical prediction to be Gaussian by applying Wick's theorem we find the following coupled equations for the density $\langle \hat{a}^\dagger\hat{a} \rangle$ and the expectation value $\langle \hat{a}^2 \rangle$ in the steady-state:
\begin{eqnarray}
0 &&= - \gamma \left(\langle \hat{a}^\dagger\hat{a} \rangle - |\alpha|^2\right) - 2 U \text{Im}\left[\alpha^{*2} \langle \hat{a}^2 \rangle \right]; \\
0 &&= 2\left( -\Delta - i\frac{\gamma}{2} \right) \left(\langle \hat{a}^2 \rangle - \alpha^2\right) + U \langle \hat{a}^2 \rangle+ 2 U \left( 3\langle \hat{a}^\dagger\hat{a} \rangle\langle \hat{a}^2 \rangle - |\alpha|^2\langle \hat{a}^2 \rangle - 2 \alpha^2\langle \hat{a}^\dagger\hat{a} \rangle \right). \nonumber
\end{eqnarray}
We have neglected the dependence of the field $\alpha$ on the quadratic correlation functions which is valid for sufficiently large photon density. 
These equations together with the expression (\ref{QuadVar}) for the variance of the quadrature lead to the Gaussian approximation presented in Fig \ref{Fig3}.

\section{Quantum trajectory simulation with photon counting measurement \label{ALG}}
We give a brief summary of the quantum trajectory algorithm with an external photon counting measurement, more details can be found in various standard textbooks such as for example Ref. \cite{breuer2002theory}. Starting from the system wavefunction $\rvert\psi(t)\rangle$ at time $t$ the quantum trajectory simulation algorithm for the evolution of the wavefunction with a discretized timestep $dt$ for a master equation of the form (\ref{eq:Master}) can be summarized as follows:
\begin{enumerate}
\item Evolve the system with the non-hermitian effective Hamiltonian $\hat{H}_{eff} = \hat{H} - i\frac{\gamma}{2}\sum_{j}\hat{a}^\dagger_{j}\hat{a}_{j}$, i.e. $\rvert\psi'(t+dt)\rangle = e^{-i \hat{H}_{eff} dt} \rvert\psi(t)\rangle$.
\item Normalize the wavefunction: $\rvert\psi(t+dt)\rangle = \rvert\psi'(t+dt)\rangle/\sqrt{\langle \psi'(t+dt)\rvert\psi'(t+dt)\rangle}$.
\item Calculate the probabilities $\{p_j\}$ for a quantum jump of the mode $j$ at time $t+dt$: $p_j = \gamma \langle \psi(t+dt)\rvert\hat{a}^\dagger_{j}\hat{a}_{j}\rvert\psi(t+dt)\rangle$.
\item Draw a random number from a uniform distribution between $0$ and $1$ and compare it with the probabilities $\{p_j\}$ to determine whether a mode performs a quantum jump.
\item If no quantum jump occurs proceed to the next time step by restarting at step 1. 
\item If mode $j$ performs a quantum jump update and normalize the wavefunction as follows: $\psi(t+dt) \to \hat{a}_{j}\psi(t+dt)/\sqrt{\langle \psi(t+dt)\rvert\hat{a}^\dagger_{j}\hat{a}_{j}\rvert\psi(t+dt)\rangle}$. Finally, proceed to the next time step by restarting at step 1.
\end{enumerate}
The steady-state density matrix can be constructed by averaging the density operator $\rvert\psi(t)\rangle \langle \psi(t) \rvert$ over time. 

This procedure does not depend on which basis is considered, i.e. whether the description is in the reciprocal space or in the real space. There is only a conceptual difference. If one considers the real space the algorithm corresponds to a measurement of the photons that leave the spatially separated resonators. For the reciprocal space the measurement counts the photons leaving the system in the reciprocal space , i.e. the photons in the bonding and anti-bonding modes. This has no influence on the average behavior of the density matrix.

\section{Gaussian approximation for the reduced density matrix of the anti-bonding mode \label{GaussDM}}
We use the Gutzwiller decoupling for the density matrix in the reciprocal space (\ref{KDC}) and truncate the Hilbert space for the anti-bonding mode to Gaussian states. The reduced density matrix $\hat{\rho}^{(AB)}$ is then completely determined by the correlation functions up to quadratic order. Since the anti-bonding mode is not externally driven the expectation value of the field is equal to zero: $\langle \hat{a}_{AB}\rangle = 0$. The relevant quadratic expectation values are thus the density $\langle \hat{a}^\dagger_{AB}\hat{a}_{AB}\rangle$ and $\langle \hat{a}_{AB}\hat{a}_{AB}\rangle$.
Applying Wick's theorem for a quadratic density matrix leads to the following coupled equations for the steady-state:
\begin{eqnarray}
0 &&=  \gamma \langle \hat{a}^\dagger_{AB}\hat{a}_{AB}\rangle + U \text{Im}\left[\langle \hat{a}^\dagger_{B}\hat{a}^\dagger_{B}\rangle\langle\hat{a}_{AB}\hat{a}_{AB}\rangle  + \langle \hat{a}^\dagger_{AB}\hat{a}^\dagger_{AB}\rangle\langle\hat{a}_{AB}\hat{a}_{AB}\rangle \right]; \\
0 &&= 2\left(-\Delta + J + U\langle \hat{a}_{B}\hat{a}_{B}\rangle  -i\frac{\gamma}{2} \right)\langle \hat{a}_{AB}\hat{a}_{AB}\rangle + \frac{U}{2} \left(2 \langle \hat{a}^\dagger_{AB}\hat{a}_{AB}\rangle + 1  \right)\langle \hat{a}_{B}\hat{a}_{B}\rangle.
\end{eqnarray}
These equations have to be solved self-consistently with the single-mode master equation for the bonding mode for which the exact result derived in Refs. \cite{minganti2016exact, PhysRevA.94.033841} can be used.

In order to calculate the distance (\ref{dist}) with respect to the exact density matrix $\hat{\rho}_{\text{ex}}$ we need the density matrix for the anti-bonding mode $\hat{\rho}_{AB}$, instead of the correlation functions. This corresponds to a squeezed thermal state:
\begin{equation}
\hat{\rho}_{AB} = \hat{S}\left(\xi\right) \hat{\rho}\left(n\right) \hat{S}\left(\xi\right)^\dagger,
\label{SqTher}
\end{equation}
with $\hat{S}\left(\xi\right)$ the squeezing operator:
\begin{equation}
\hat{S}\left(\xi\right) = e^{\left(\xi \hat{a}^{\dagger 2} - \xi^* \hat{a}^2 \right)/2}
\end{equation}
and $\hat{\rho}\left(n\right)$ the thermal density matrix with average photon number $n$:
\begin{equation}
\hat{\rho}\left(n\right) = \frac{1}{1+n} \sum_{m=0}^\infty \left( \frac{n}{1+n} \right)^m \rvert m \rangle \langle m \rvert ,
\end{equation}
where $\rvert m \rangle$ denote a number Fock state. If we write the squeezing parameter as $\xi = te^{i\theta}$ we find that a state with expectation values $\langle \hat{a}^\dagger_{AB}\hat{a}_{AB}\rangle$ and $\langle \hat{a}_{AB}\hat{a}_{AB}\rangle$ corresponds to the density matrix (\ref{SqTher}) with:
\begin{eqnarray}
\theta =& \arg\left[\langle \hat{a}_{AB}\hat{a}_{AB}\rangle \right] ; \\
r =& \ln\left[\frac{1 + 2\langle \hat{a}^\dagger_{AB}\hat{a}_{AB}\rangle + 2 \rvert\langle \hat{a}_{AB}\hat{a}_{AB}\rangle\rvert}{1 + 2\langle \hat{a}^\dagger_{AB}\hat{a}_{AB}\rangle - 2\rvert\langle \hat{a}_{AB}\hat{a}_{AB}\rangle\rvert}   \right]/4; \\
n =& \frac{1}{2}\left(\sqrt{\left(1 + 2\langle \hat{a}^\dagger_{AB}\hat{a}_{AB}\rangle \right)^2 -4 \rvert \langle \hat{a}_{AB}\hat{a}_{AB}\rangle\rvert^2} - 1\right).
\end{eqnarray}
With $\arg[x]$ the function that takes the argument of a complex number $x$ and $\ln[x]$ the natural logarithm of $x$. This allows us to construct the reduced density matrix for the Gaussian anti-bonding mode and calculate the distance (\ref{dist}) with respect to the exact result, as presented in Fig. \ref{Fig2}.

\section{Gaussian approximation for the anti-bonding mode wavefunction during a quantum trajectory simulation \label{GaussWF}}
We discuss how the Gaussian approximation is performed for the wavefunction of the anti-bonding mode during a quantum trajectory simulation. The Gutzwiller Ansatz with respect to the reciprocal space leads to the following effective Hamiltonian for the anti-bonding mode:
\begin{eqnarray}
\hat{H}_{eff}^{(AB)} = &&\left(-\Delta + J + U \langle \hat{a}^\dagger_{B}\hat{a}_{B}\rangle - i\frac{\gamma}{2}\right) \hat{a}^\dagger_{AB}\hat{a}_{AB} \nonumber\\
&&+ \frac{U}{4}\left(\hat{a}^\dagger_{AB}\hat{a}^\dagger_{AB}\hat{a}_{AB}\hat{a}_{AB} + \langle\hat{a}^\dagger_{B}\hat{a}^\dagger_{B}\rangle\hat{a}_{AB}\hat{a}_{AB} + \hat{a}^\dagger_{AB}\hat{a}^\dagger_{AB}\langle\hat{a}_{B}\hat{a}_{B}\rangle \right)
\label{EffHamAB}
\end{eqnarray}
The bonding mode expectation values are obtained from a parallel quantum trajectory simulation for this mode for which the full Hilbert space is considered. 

A Gaussian state is completely determined by its correlation functions up to quadratic order. From the effective Hamiltonian (\ref{EffHamAB}) we determine the  equations of motion for these correlation functions. These are needed to implement step 1 of the algorithm outlined in appendix \ref{ALG} and are given by:
\begin{eqnarray}
i \partial_t \langle \hat{a}_{AB}\rangle &&= \left(-\Delta + J + U \langle \hat{a}^\dagger_{B}\hat{a}_{B}\rangle - i\frac{\gamma}{2}\right)\langle \hat{a}_{AB}\rangle + \left(\frac{U}{2} - i\gamma\right)\langle \hat{a}_{AB}^\dagger\hat{a}_{AB}\hat{a}_{AB}\rangle + \frac{U}{2}\langle \hat{a}_{AB}^\dagger\rangle \langle\hat{a}_{B}\hat{a}_{B}\rangle; \nonumber \\
i \partial_t \langle \hat{a}_{AB}\hat{a}_{AB}\rangle &&= 2\left(-\Delta + J + U \langle \hat{a}^\dagger_{B}\hat{a}_{B}\rangle -\frac{U}{2} - i\frac{\gamma}{2}\right)\langle \hat{a}_{AB}\hat{a}_{AB}\rangle + \left(U - i\gamma\right)\langle \hat{a}_{AB}^\dagger\hat{a}_{AB}\hat{a}_{AB}\hat{a}_{AB}\rangle \nonumber\\
&&+ \frac{U}{2}\left( 2\langle \hat{a}_{AB}^\dagger \hat{a}_{AB}\rangle + \langle \psi \rvert \psi \rangle \right)\langle\hat{a}_{B}\hat{a}_{B}\rangle; \nonumber \\
i \partial_t \langle \hat{a}_{AB}^\dagger\hat{a}_{AB}\rangle &&=  -i\gamma\left(\langle \hat{a}_{AB}^\dagger\hat{a}_{AB}\rangle + \langle \hat{a}_{AB}^\dagger\hat{a}_{AB}^\dagger\hat{a}_{AB}\hat{a}_{AB}\rangle \right) + \frac{U}{2}\left(\langle \hat{a}_{AB}^\dagger \hat{a}_{AB}^\dagger\rangle \langle\hat{a}_{B}\hat{a}_{B}\rangle- \langle \hat{a}_{AB}\hat{a}_{AB}\rangle \langle\hat{a}_{B}^\dagger\hat{a}_{B}^\dagger\rangle \right); \nonumber \\
i \partial_t\langle \psi \rvert \psi \rangle && = - i \gamma \langle \hat{a}_{AB}^\dagger\hat{a}_{AB}\rangle.
\label{DetTime}
\end{eqnarray}
Since the Hamiltonian (\ref{EffHamAB}) is not Hermitian the norm of the wavefunction $\rvert \psi \rangle$ is not conserved during the evolution. We assume the wavefunction to be Gaussian by applying Wick's theorem for the higher order correlation functions:
\begin{eqnarray}
\langle \hat{a}_{AB}^\dagger\hat{a}_{AB}\hat{a}_{AB}\hat{a}_{AB}\rangle &&= 3 \frac{\langle \hat{a}_{AB}^\dagger\hat{a}_{AB}\rangle\langle\hat{a}_{AB}\hat{a}_{AB}\rangle}{\langle \psi \rvert \psi \rangle} - 2 \frac{\langle \hat{a}_{AB}^\dagger \rangle\langle\hat{a}_{AB}\rangle^3}{\langle \psi \rvert \psi \rangle^3}  \nonumber\\
\langle \hat{a}_{AB}^\dagger\hat{a}_{AB}\hat{a}_{AB}\rangle &&= \frac{\langle \hat{a}_{AB}^\dagger\rangle\langle\hat{a}_{AB}\hat{a}_{AB}\rangle}{\langle \psi \rvert \psi \rangle} + 2 \frac{\langle \hat{a}_{AB}^\dagger\hat{a}_{AB}\rangle\langle\hat{a}_{AB}\rangle}{\langle \psi \rvert \psi \rangle} - 2\frac{\langle \hat{a}_{AB}^\dagger\rangle\langle\hat{a}_{AB}\rangle^2}{\langle \psi \rvert \psi \rangle^2} \nonumber\\
\langle \hat{a}_{AB}^\dagger\hat{a}_{AB}^\dagger\hat{a}_{AB}\hat{a}_{AB}\rangle &&= 2\frac{\langle\hat{a}_{AB}^\dagger\hat{a}_{AB}\rangle^2}{\langle \psi \rvert \psi \rangle} + \frac{\rvert \langle\hat{a}_{AB}\hat{a}_{AB}\rangle \rvert^2}{\langle \psi \rvert \psi \rangle} - 2 \frac{\rvert\langle\hat{a}_{AB}\rangle \rvert^4}{\langle \psi \rvert \psi \rangle^3} 
\label{Wick}
\end{eqnarray}
Again we had to keep track of the norm of the wavefunction.
The equations (\ref{DetTime}) together with the Wick decoupling (\ref{Wick}) allow to perform the deterministic time evolution of the correlation functions, corresponding to step 1 of the algorithm summarized in appendix \ref{ALG}. 

The expectation value $\langle \hat{a}_{AB}^\dagger\hat{a}_{AB}\rangle$ determines the probability for a quantum jump. Whether a quantum jump occurs is determined stochastically by comparing a random number to $\gamma\langle \hat{a}_{AB}^\dagger\hat{a}_{AB}\rangle$. If this occurs the correlation functions are updated as follows (the wavefunction is now normalized):
\begin{eqnarray}
\langle \hat{a}_{AB}\rangle &&\to \frac{\langle \hat{a}_{AB}^\dagger\hat{a}_{AB}\hat{a}_{AB}\rangle}{\langle \hat{a}_{AB}^\dagger\hat{a}_{AB}\rangle} = \frac{\langle \hat{a}_{AB}^\dagger\rangle\langle\hat{a}_{AB}\hat{a}_{AB}\rangle + 2 \langle \hat{a}_{AB}^\dagger\hat{a}_{AB}\rangle\langle\hat{a}_{AB}\rangle - 2\langle \hat{a}_{AB}^\dagger\rangle\langle\hat{a}_{AB}\rangle^2}{\langle \hat{a}_{AB}^\dagger\hat{a}_{AB}\rangle} \nonumber \\
\langle \hat{a}_{AB}\hat{a}_{AB}\rangle &&\to \frac{\langle \hat{a}_{AB}^\dagger\hat{a}_{AB}\hat{a}_{AB}\hat{a}_{AB}\rangle}{\langle \hat{a}_{AB}^\dagger\hat{a}_{AB}\rangle} = \frac{3 \langle \hat{a}_{AB}^\dagger\hat{a}_{AB}\rangle\langle\hat{a}_{AB}\hat{a}_{AB}\rangle - 2 \langle \hat{a}_{AB}^\dagger \rangle\langle\hat{a}_{AB}\rangle^3}{\langle \hat{a}_{AB}^\dagger\hat{a}_{AB}\rangle} \nonumber \\
\langle \hat{a}_{AB}^\dagger\hat{a}_{AB}\rangle &&\to \frac{\langle \hat{a}_{AB}^\dagger\hat{a}_{AB}^\dagger\hat{a}_{AB}\hat{a}_{AB}\rangle}{\langle \hat{a}_{AB}^\dagger\hat{a}_{AB}\rangle} = \frac{2\langle \hat{a}_{AB}^\dagger\hat{a}_{AB}\rangle^2+ \rvert\langle\hat{a}_{AB}\hat{a}_{AB}\rangle\rvert^2 - 2 \langle\hat{a}_{AB}^\dagger\rangle^2\langle\hat{a}_{AB}\rangle^2}{\langle \hat{a}_{AB}^\dagger\hat{a}_{AB}\rangle} \nonumber \\
\end{eqnarray} 
where we again used Wick's theorem.

In order to construct the density matrix we need the wavefunction $\rvert \psi\rangle$ associated to these correlation functions. A general quadratic wavefunction $\rvert \psi\rangle$ corresponds to a squeezed and displaced vacuum $\rvert 0\rangle$:
\begin{equation}
\rvert \psi\rangle = \hat{D}\left(\alpha \right)\hat{S}\left(\xi\right) \rvert 0\rangle,
\label{WF}
\end{equation}
with $\hat{D}\left(\alpha\right)$ the displacement operator:
\begin{equation}
\hat{D}\left(\alpha\right) = e^{\alpha \hat{a}^{\dagger} - \alpha^* \hat{a}}
\end{equation}
and $\hat{S}\left(\xi\right)$ the squeezing operator:
\begin{equation}
\hat{S}\left(\xi\right) = e^{\left(\xi \hat{a}^{\dagger 2} - \xi^* \hat{a}^2 \right)/2}.
\end{equation}
If we write the sqeezing parameter as $\xi = re^{i\theta}$ the following expressions give the wavefunction parameters in terms of the correlation functions \cite{gerry2005introductory}:
\begin{eqnarray}
\alpha =&& \langle \hat{a}_{AB}\rangle; \\
r =&& \text{arcsinh}\left[\sqrt{\langle \hat{a}_{AB}^\dagger\hat{a}_{AB}\rangle - \rvert\langle \hat{a}_{AB}\rangle\rvert^2} \right] \\
\theta =&& -i \ln \left[\frac{\rvert\langle \hat{a}_{AB}\rangle\rvert^2 - \langle \hat{a}_{AB}\hat{a}_{AB}\rangle}{\sqrt{\langle \hat{a}_{AB}^\dagger\hat{a}_{AB}\rangle - \rvert\langle \hat{a}_{AB}\rangle\rvert^2}\sqrt{1 + \langle \hat{a}_{AB}^\dagger\hat{a}_{AB}\rangle - \rvert\langle \hat{a}_{AB}\rangle\rvert^2}} \right].
\end{eqnarray}
This allows us to construct the correponding density matrix by averaging over the density operator $\rvert \psi\rangle \langle \psi\rvert$.

\section{Gaussian approximation for the driven-dissipative dimer \label{DDDGauss}}
We now discuss the Gaussian approximation for the fluctuations around the semiclassical prediction for the driven-dissipative Bose-Hubbard dimer. We only consider the homogeneous solutions for which the anti-bonding mode is unoccupied: $\langle \hat{a}_{AB} \rangle = 0$. The semiclassical steady-state prediction for the field of the bonding mode $\alpha_B = \langle \hat{a}_B \rangle$ is then determined by the following equation:
\begin{equation}
\left(-\Delta - J - i\frac{\gamma}{2} + \frac{U}{2} |\alpha_B|^2 \right)\alpha_B + \sqrt{2}F = 0.
\end{equation}
This equation leads to expression (\ref{eq:SCDen}) for the local density $n = |\alpha_B|^2/2$. Assuming the fluctuations around the semiclassical field to be quadratic by applying Wick's theorem leads to the following coupled equations for the steady-state:
\begin{eqnarray}
0 =&&  \gamma \left(\langle \hat{a}^\dagger_{B}\hat{a}_{B} \rangle -  |\alpha_B|^2  \right)+ U \text{Im}\left[\left(\langle \hat{a}_{AB}\hat{a}_{AB} \rangle + \langle \hat{a}_{B}\hat{a}_{B} \rangle\right)^*\left( \langle \hat{a}_{B}\hat{a}_{B} \rangle - \alpha_B^{2} \right) \right]; \nonumber\\
0 =&&  \gamma \langle \hat{a}^\dagger_{AB}\hat{a}_{AB} \rangle + U \text{Im}\left[\left(\langle \hat{a}_{AB}\hat{a}_{AB} \rangle + \langle \hat{a}_{B}\hat{a}_{B} \rangle\right)^*\langle \hat{a}_{AB}\hat{a}_{AB} \rangle \right]; \nonumber\\
0 =&& 2 \left[-\Delta - J + U\left(\langle \hat{a}^\dagger_{B}\hat{a}_{B} \rangle + \langle \hat{a}^\dagger_{AB}\hat{a}_{AB} \rangle \right) -i\frac{\gamma}{2}  \right]\left(\langle \hat{a}^\dagger_{B}\hat{a}_{B} \rangle -  |\alpha_B|^2  \right) \nonumber \\
 &&+ \frac{U}{2} \left(\langle \hat{a}_{AB}\hat{a}_{AB} \rangle + \langle \hat{a}_{B}\hat{a}_{B} \rangle\right)\left(1 + 2 \langle\hat{a}^\dagger_{B}\hat{a}_{B} \rangle -  2|\alpha_B|^2 \right); \nonumber \\
0 =&& 2 \left[-\Delta + J + U\left(\langle \hat{a}^\dagger_{B}\hat{a}_{B} \rangle + \langle \hat{a}^\dagger_{AB}\hat{a}_{AB} \rangle \right) -i\frac{\gamma}{2}  \right] \langle \hat{a}^\dagger_{AB}\hat{a}_{AB} \rangle \nonumber \\
&&+ \frac{U}{2} \left( \langle \hat{a}_{AB}\hat{a}_{AB} \rangle + \langle \hat{a}_{B}\hat{a}_{B} \rangle\right)\left(1 + 2 \langle \hat{a}^\dagger_{AB}\hat{a}_{AB} \rangle \right).
\end{eqnarray}
These equations allow to determine the variances of the quadratures of the bonding and the anti-bonding modes and the variances of the two EPR-like operators (\ref{EPR}) as presented in the lower panel of Fig. \ref{Fig4}.

\bibliography{manusc}

\end{document}